# When Everything hinges *On Gravitation*

*Karin Verelst*[1]
kverelst@vub.ac.be

*Materiam coelorum fluidam esse* — *Isaac Newton (1682)*
*Spatium est entis quatenus ens affectio* — *Isaac Newton (1683)*

**Abstract**

*Newton's basic ideas developed and evolved throughout his career and changed in sometimes surprisingly profound ways. In this paper I propose an outline of the evolution of Newton's conceptual framework by following the development of his ideas throughout the early work preceding the first edition of the* Principia*, and thus to complete the work that has been done by Whiteside and Ruffner with respect to Newton's mechanics. I shall show that these evolutions — the mechanical and the metaphysical — are closely interrelated. My focus will be on a key text that marks a turning point both from the metaphysical and the methodological point of view: the* "De gravitatione et aequipondio fluidorum". *Rather than looking upon it as an isolated fact, I establish the connections of this text to other manuscripts from the same period, primarily the manuscript "Elements of Mechanicks" in the Hall & Hall edition, as well as the two variants of the "De Motu",* to which it can be said to relate as to a "zero release".

**Introduction**

It is known that the period of Newton's lifetime between 1679 and 1694 was filled with intellectual transition and turmoil. Nevertheless, relatively little work has been done on explicit attempts to outline the specific paths and turns followed by his conceptual development throughout this same period.[2] Indeed, a

---

[1] FUND-CLEA, Dept. of Mathematics, Vrije Universiteit Brussel and Open Universiteit Nederland, Faculty of Cultural Studies.

[2] With regard to his mechanics D. T. Whiteside, "Before the *Principia*," *Journal for the History of Astronomy*, **1,** 1970, pp. 5–19, and J.A. Ruffner, "Newton's Propositions on Comets: Steps in Transition, 1681–84", *Archive for the History of Exact Science*, **54**, 2000, pp. 259–277. J.W. Herrivel's book, although critically assessed by Whiteside, remains the source for

certain bias towards an "ahistorical" reading of Newton's works, especially those concerning natural philosophy, continues to prevail. This may be somewhat of a surprise in the light of the overwhelming richness of the material that allows for a more critical approach:

> *The conventional view of the prehistory of Newton's synthesis in the Principia of his predecessors' work in planetary theory and terrestrial gravitation is still not seriously changed from that which Newton himself chose to impose on his contemporaries at the end of his life.*[3]

Simon Schaffer has discussed at length and in depth the origins of biases of this kind, and the traps which they lay out on the way of the historian or philosopher engaged in the exercise of "Newtonian hermeneutics": a different reading is not neccesarily a wrong one. *Interpretation is an accomplishment and this accomplishment is the historian's explanandum*.[4] I happily second that, but add the caveat that some interpretations might nevertheless be better than other

---

young Newton's attempts in dynamics: *The background to Newton's Principia. A study of Newton's dynamical researches in the years 1664-1684, based on original manuscripts from the Portsmouth Collection in the Library of the University of Cambridge*, Oxford University Press, Oxford, 1965. The only systematic treatment that links practical work in science to developing metaphysical ideas with Newton is B.J.T. Dobbs, *The Janus Face of Genius*, Cambridge University Press, Cambridge, 1990. On the other hand, a vivid interest in the relation between Newton's theology and his natural philosophy exists. As a guide through this literature, the reader can start with S. Snobelen, "'To discourse of God': Isaac Newton's heterodox theology and his natural philosophy", in P.B. Wood (ed.), *Science and dissent in England*, 1688-1945, Aldershot: Ashgate, 2004, pp. 39-65.

[3] D.T. Whiteside, "Newton's Early Thoughts on Planetary Motion: a Fresh Look", *The British Journal for the History of Science*, **2,** 6, 1964, pp. 118-137.

[4] S. Schaffer, "Comets & Idols: Newton's cosmology and Political Theology", in: P. Theerman and A.F. Seeff, *Action and reaction. Proceedings of a Symposium to Commemorate the Tercentenary of Newton's* Principia, University of Delaware Press, Newark, 1993.

ones, especially those which do not only take the different audiences addressed[5], but also the different stages of attestable conceptual development of the author into account, his own intellectual history, so to say. In that context, issues of dating and interrelation of manuscripts and published works take on another meaning than merely that of strategies in an interpretative debate. For this reason an editorial method has been established long ago which gives the researcher access to comparable text sources, provided with a critical apparatus. Saidly enough, no such critical editions exist in the field of Newton studies up to to-day, probably testifying to the fact that this field is even more infected by the disease of "hermeneutical essentialism" than other, related ones, say, the study of authors from the same period, like Galilei, Descartes, Huygens, or Leibniz.

Symptomatic for this situation — as Schaffer himself points out[6] — are the widely diverging dates ascribed to a text which is clearly of key importance, the *De Gravitatione et aequipondio fluidorum (On Gravity and the aequilibrium of fluids)*. This text, its title notwithstanding, deals first and foremost with a thorough criticism of the metaphysical positions elaborated by Descartes in his *Principia Philosophiae*.[7] Hall & Hall, who published the DG text with a translation[8], as well as McGuire in his edition of an equally enigmatic manuscript[9], qualify it as "youthful" (around 1670), while Janiak in his new translation situates it around 1680.[10] Recently, Edward Richards even argued that the DG is to be placed after the *Principia*![11] My aim in this paper is to show that Newton wrote the *De Gravitatione* in the same period he was working on the comets during the years 1681-1684, and more specifically around 1682-1683. I furthermore establish the close links it entertains both by content and method to the different *De Motu* drafts and to the first published edition of the *Principia* (1687). The fact that Newton's ideas on cometary motion underwent drastic changes in the period of the *De motu* drafts has been thematised again by Schaffer in the second part of the paper quoted above. His perspective, however, is the theologically steered cosmology Newton starts to develop during that same period and the active rôle comets play therein. Schaffer hints at the physical import of this theological scheme, but places its elaboration much later and on a more speculative level.[12] We shall see how the question of comets directly leads to the first edition of the *Principia*, because it made unavoidable the fundamental rejection of both the physics and the metaphysics of Descartes.

---

[5] More recent work on this issue has been done by R. Iliffe, e.g., "Abstract considerations: disciplines, audiences and the incoherence of Newton's natural philosophy", *Studies in the History and Philosophy of Science*, March 2003.
[6] S. Schaffer, *o.c.,* p. 208.
[7] R. Descartes, "Principia Philosophiae", in *Oeuvres complètes de R. Descartes,* Tome VIII-I, C. Adam et P. Tannery (eds*.), * Cerf, Paris, 1897-1913.
[8] A. Rupert Hall and M. Boas Hall (eds.), *Unpublished Scientific Papers of Isaac Newton*, Cambridge University Press, Cambridge, 1962.

---

[9] Following usage, I shall label it after its first words *Tempus et Locus*. See McGuire, "Newton on Place, Time and God: an Unpublished source", *The British Journal for the History of science*, **11**, 38, 1978, pp. 114-129.

[10] I. Newton, *Philosophical Writings,* A. Janiak (ed.), Cambridge University Press, Cambridge, 2004.

[11] E.T. Richards, *A Philosophical Analysis of Newton's Arguments against Cartesianism as found in 'De Gravitatione'*, Doctoral Dissertation, Boston University, 2006.

[12] *Because Newton made comets' tails the bearers of celestial activity, he had to explain how they worked in void space. Both Gregory and Henry Pemberton, editor of the third edition of the* Principia*, found trouble making sense of this thought. (...) Using his alchemical and matter theoretic drafts of the 1670s, reworking the initial Definitions to the* Principia *on transmutation of matter, and adding queries to successive editions of the* Opticks *after 1704, Newton spelled out the various ways in which the activity of the cosmos could be sustained by the rare fluids carried by comets.* We shall see below how this impinges directly on another important shift in Newton's ideas on methodology which will follow soon after the publication of PR I, in the early 1690s. For the quote, see S. Schaffer, o.c., pp. 218-219.

I thus want to contribute something to the outline of a comprehensive map of evolution of Newton's views up to the advent of the first *Principia* by studying the contextus of this text in some detail. It will appear that Cohen's advise is applicable more generally when it comes to the relation between "science" and "philosophy" in Newton's early years:

> *(..) to understand Newton's philosophy of science, one must not characterise his early and most creative periods of scientific thought by later slogans such as "Hypotheses non fingo". Rather we must see Newton's thought in its development as he progressed from a tolerance of certain types of hypotheses, **especially speculations as to the cause of phenomena**, to an alledged abhorrence of them all.*[13]

The *De Gravitatione* played a pivotal rôle in the decisive metaphysical shift Newton made away from Cartesianism shortly before he set out to write his *Principia*. It also bears witness to his taking the first, still hestitant, steps towards what will become only much later, in the second and third edition of the *Principia*, his celebrated *Philosophia experimentalis*.

**Some textual interrelationships**

The ahistorical bias metioned above is obviously what led the Halls and McGuire into believing that the DG is a youthful work belonging to his "student days"[14], while this is plainly impossible when placed

---

[13] I.B. Cohen, "Hypotheses in Newton's Philosophy", *Physis*, VIII, 1966, p. 163.
[14] Hall & Hall, p. 75.

side by side to a text of which the early origin is indubitable, e.g. the *Lawes of Motion*. Hall & Hall state (on p. 76 of their introduction): *Both this short paper and n° 2,* 'De gravitatione et aequipondio fluidorum' *appear to spring from Newton's critical reaction to the* Principia Philosophiae *of Descartes.* And this may well be the case, for we know with certainty that Newton read Descartes's *Principia* already as a student[15], but it does not therefore follow that they have been written in the same period, with the same intention, nor necessarily in an early phase of his lifetime. This is nevertheless their point with regard to them: *Nos. 1 and 2 clearly antidate the* Principia *by many years* (p. 75). There are a few interesting similarities between both these tracts and the *Principia*, like the characterisation of the motion of bodies as *their passing out of one place or part of space into another, through all ye intermediate space is their motion*.[16] There also is the link of force to the moving body's "bulk" (in itself an early, imprecise precursor of his later concept of "mass"). But the criticism of Descartes in *Lawes* is implicit, and focused on the mathematical description of the "rules of Reflection" for "absolutely hard" bodies. There is no trace of a rejection of the spatial aether[17], on the

---

[15] J.E. McGuire and M. Tamny, Certain Philosophical Questions. Newton's Trinity Notebook [TNB in what follows], Cambridge University Press, Cambridge, 2002 [1983], p. 127 sq. "The Cartesian Influence").

[16] *Lawes*, Hall & Halll, p. 157.

[17] The statement at the beginning of the first paragraph ("There is a uniform extension, are expansion continued every way without bounds: in which all bodies are, each in severall parts of it") cannot suffice for making such a claim. As McGuire and Tamny remark in their commentary on the Trinity Notebook, a collection of even earlier texts: *There is more, of course, to a theory of absolute space than the conception of void and penetrable expanses that lack resistance to motion. To be sure, these features will characterise any conception of space that can be said to 'contain' the movements of corporeal existents. But more then this**, the categorical nature of space must be considered.*** *Is it a substance or an accident? Or does its mode of existence differ from the traditional categories of being? What is the connection determined relative to the senses and motions with respect to insensible space itself?* They add in ft.

contrary, it is stated unambigiously that *[bodies'] motions are continually impeded by ye mediums in which they move*[18], while the criticism of Descartes in de DG is explicit and aimed at the very foundations of his metaphysical framework. This profound difference is matched by an equally profound difference in style. On the one hand the schoolish manner in which the tract on 'Lawes' is composed, with *quaestiones* to solve ("Now to know the real quantity..."; "now... to know how these bodies shall be reflected") in the abbreviated, formalistic style which still was prevalent in university education in Newton's student days.[19] There is nothing in the DG that matches this stylistic mark to any comparable extend. Moreover the hesitant, unprecise phrasing of basic concepts of metaphysical import in *Lawes*: *[bodies] have a relenting softenesse & springynesse which makes their contact be for some time & in more points then one*[20] as contrasted to the sec and self-assured formulations ("Hactenus de natura corporea"[21]) on, e.g., the impenetrability of matter in the DG. The point for Newton in *Lawes* is that Descartes's theory of how objects move through a medium — their percussions and reflections are therefore involved — does not hold on purely mathematical grounds (Newton succeeds for the first time in the construction of the parallelogram of forces). But *Lawes* still completely thinks in terms of an aether. *De Gravitatione,* apart from positing right from the start that mathematical philosophy is superior to mere natural philosophy, sets out to

---

213: *These questions are systematically considered in* De gravitatione. McGuire & Tamny, *o.c.,* p. 124 (my bold).
[18] Hall & Hall, p. 163.
[19] And even went through a revival at Cambridge. Cfr. S. Mandelbrote's remarks on the restauration of the Divinity Act by the mid 1670s Regius Professor J. Beaumont, Master of Peterhouse, during a lecture "Becoming heterodox in seventeenth-century Cambridge: the case of Isaac Newton", held at a workshop at the Belgian Royal Academy in Brussels, september 2007.
[20] Hall & Hall, p. 163.
[21] Hall & Hall, p. 111.

develop a sound fluid mechanics as a basis for a "system of the world", and then turns — at first sight a contradiction between the formal and the metaphysical content of the DG — to the systematic development of the notion of a physical vacuum into the metaphysical absolute space which functions as the global system of reference, as opposed to the local one of relative motion of bodies referring to themselves only:

> *Corporum positiones, distantiae, et motus locales ad spatij partes referenda sunt. (...) Denique spatium est eternae durationis et immutabilis naturae, idque quod sit aeternus et immutabilis entis effectus emanativus.[22]*

There is no real contradiction, however: Newton comes to realise fully and completely the untenability of Descartes's metaphysical positions only when he tries to work them out into a sound mathematical theory that is able to decribe facts — like, as we will see, cometary motions — hence the critcism of Descartes's *Principia* in the main part of DG. The *Tempus and Motus* texts indeed relate to this; they take up and refine the arguments of DG, as McGuire reckons and elaborates in detail, but because of his dating of DG (between 1666-1670) it is impossible for him to appreciate the change Newton's ideas underwent.

> *Significantly, the manuscript (Tempus et Locus) is also based on* De gravitatione. *This important treatise was written somewhere between 1666 and 1670. It constitutes an important set of criticisms of Descartes (...) Moreover, it represents Newton's most philosophical account of his doctrines of absolute motion, place,*

---

[22] Hall & Hall, p. 104.

*space and time. At every turn, Newton's characteritic theories are developed in opposition to Descartes'.*[23]

One wonders how the distinguished editor of both this text and of the Trinity Notebook with its references to the 1675 Oldenburg letter and the 1682 *de Cometis* ms. ULC 3965 (see further) could have failed to see the apparent contradiction. I believe this is a consequence of the bias according to which Newton did not have to go through any real intellectual evolution; he was born destined to destroy the Cartesian fantasies from the start. Indeed, the above contradiction dissappears entirely once you see that the only thing wrong is the date ascribed to DG. From the moment it is accepted that DG must have been written somewhere around 1683, everything falls into place, and the ensuing statement obtains its full impact:

*Without the stimulus of Descartes' philosophy it is doubtful whether Newton's thinking would have found the kind of expression which is revealed in De gravitatione. Nor is this the end of the Frenchman's influence: a careful comparison of the structure of the scholium on space and time with De gravitatione shows that it is based on the latter.*[24]

**The need for a reading of the *De Gravitatione* as a whole**

I believe that Newton's idea about the system of the world up to then had remained "Cartesian" in the sense that it was based on *the*

---

[23] McGuire, *Tempus et Locus*, p. 124. *The Tempus et Locus* fragments themselves fit in the post-*Principia* period, probably from 1695 onwards, when he was drafting his first versions of the *General Scholium*. This latter text features remarkable resemblances to the DG in many respects, especially when its drafts are taken into account. For SG's prehistory, see S. Ducheyne, "The General Scholium: Some Notes on Newton's Published and Unpublished Endeavours", *Lias: Sources and Documents Relating to the Early Modern History of Ideas*, **33**, 3, 2006, pp. 223-274.

[24] McGuire, Tempus et Locus, p. 124.

*aethers in the vortices of the Sun and Planets*.[25] As late as 1682, he can write *materium coelorum fluidam esse. Materiam coelorum circa centrum systematis cosmici secundum cursam planetarum gyrare*.[26] His critical attitude towards Cartesianism – already evident in de early Trinity Notebook – does not yet affect his adherence to its theoretical foundation. On the contrary, in the *DG* he initially sets out to develop a sound mathematical basis for the mechanical theory of motion based on it, the behaviour of bodies in a fluid medium, fostered — I claim — by his studies of the comets.[27] The *De Gravitatione* plainly testifies to Newton's struggle to get fluid mechanics right and his failure, which forces him into an in-depth and detailed re-examination, and ultimately rejection, of its theoretical foundations as laid out in a scholarly manner by Descartes in his *Principia Philosophiae*. This point was already made by Biarnais in her 1985 translation:

> *A cet égard, Le De Gravitatione et et Aequipondio fluidorum représente le point culminant de ces remarques, en tant que réflexion méthodique sur les fondéments de la science mécanique (...) Pourquoi enfin "science de l'équilibre des fluides et des solides dans les fluides"? Car, précisement, le Système du monde en honneur à l'époque bien que fort discuté est la théorie cartésienne*

---

[25] Letter to Oldenburg, 1675, reprinted in H. W. Turnbull, J. F. Scott, A. R. Hall, and L. Tilling (eds.), *The Correspondence of Isaac Newton*, ed. 7 vols., Cambridge University Press, Cambridge, 1959–1984., vol. I, p. 368.

[26] ULC Add. 3965.14, fl. 613r. This Ms. has been published in facsimile and transcribed with translation by Ruffner in his already cited paper, as *Propositiones de Cometis*, pp. 260-263. The statement is also quoted in *TNB*, p. 169 ft. 120.

[27] The link to the comets I owe to Ruffner's paper: *Isaac Newton's closest approach to a system of the world in the critical period 1681–84 is provided in a set of untitled propositions concerning comets*, p. 259.

> *des tourbillons où il est fait appel aux propirétés des fluides pour rendre compte du transport des planètes dans l'univers.*[28]

It did not make its way into mainstream scholarship, however, as can be inferred from the fact that Newton's manuscript is chopped into "philosophical" and "physical" pieces by commentators and translators until to-day.[29] His systematic discussion of Descartes's *Principia* is far from a mere *digression, leading very far from the announced subject of hydrostatics*[30], even though Newton himself calls it a "digression" towards the end.[31] This qualification indeed testifies for the fact that he himself had not planned to go there from the start, for he had not yet fully swallowed the consequences of his own theoretical discoveries. It wouldn't take long, however, before he understood that a farewell to Descartes also meant a farewell to fluid dynamics as a basis for a system of the world:

> *Newton clearly intended to write an elaborate treatise on hydrostatics; but, after completing a long criticism of Descartes, he seems to have lost interest in his original purpose.*[32]

Again, why would that be? Because a metaphysics based on empty space requires a mechanics that operates *in vacuo*. His abandoning of the hydrostatic project was thus far more than a mere *caprice*.

---

[28] I. Newton, *De la gravitation, ou les fondements de la mécanique classique*, M.-F. Biarnais (ed.), Les Belles Lettres, Paris, 1985, Introduction, pp. 9-10.
[29] The extend to which this approach is unfruitful becomes clear when one keeps the example of the General Scholium in mind. Cfr. De Smet, R., Verelst, K., "Newton's Scholium Generale. The Platonic and Stoic legacy: Philo, Justus Lipsius and the Cambridge Platonists", *History of Science*, xxxix, 2001.
[30] So Hall & Hall in their introduction, p. 89.
[31] Hall & Hall, p. 114.
[32] Hall & Hall, p. 76.

Interestingly enough, this is confirmed by the fact that Newton's criticism in DG does not start with Cartesian mechanism. The set of definitions exposed at the start by no means imply the vacuum, let alone absolute space. He comes to this point only after having critically discussed the Scholastic approach to body as as substance with sensible qualities on *methodological* grounds.[33] The methodological stance Newton occupies in the *De Gravitatione* is interesting, but equally ambiguous. In the introducing paragraphs, he clearly sees mathematisation as an alternative to dialectics as the "pure science" by which truths can be deduced from first principles. This he takes directly from Descartes, whose rupture with the ancient *ars inveniendi* consists precisely in the replacement of logic by mathematics.[34] The truly novel element here is that mathematics *may be applied to making clear many of the phenomena of natural philosophy*.[35] Experiment comes in not only to illustrate but, by demonstrating their applicability, to confirm the certainty of the mathematical principles used. There is still a long way to go to Newton's later empiricism! The difference between natural phlosophy and mathematics is codified stylistically: experimental results are discussed in *scholia*, a mode of presentation proper to the more informal realm of natural philosophy, while the deductive part of Newton's scheme relies on mathematical rigour. This complies to some points raised by Smith in his analysis of the methodology of the first *Principia*: there (in the Introduction) a rare but clear statement on the key rôle of empirical data ("phaenomena") is made, while at the same time mathematics is said to enhance the possibility to *argue more securely* in natural philosophy.[36] The full transition to experimental philosophy in the true sense of the word will take place only afterwards, between the first and the second edition of the *Principia*, although its *Ansatz* is already present in the *De Gravitatione*.[37] Thus, from the methodological point of view, PR I occupies an intermediate position between DG and the later PR-editions, as it does in many more respects, as shall be discussed below.

---

[33] A point that does not come out very well in the translation proposed by Janiak, pp. 12-14.
[34] *It is arithmetic and geometry, the two subjects of pure mathematics, that Descartes extols instead of dialectic or logic*. C. Sasaki, *Descartes's Mathematical Thought*, Kluwer, Dordrecht, 2003, p. 178. Newton could have learned this from Desacrtes's *Principia*, as well as from his mathematical works.
[35] In Janiak's translation, p. 12.

**A cluster of structurally related pre-Principia manuscripts**

Newton had been incited by the correspondence with Hooke in 1679 (as he himself recognises in a letter to Halley) to resume work on the problem of planetary motion.[38] At that moment he continued to adhere to the idea of the heavenly fluids as the cause of gravitaional force. What might have brought him to change his fundamental stance on his version of the Cartesian aether? I think Ruffner provides a key to the answer in his commentary on a most intriguing manuscript dealing with a subject that had attracted Newton's curiosity already at the beginning of his scientific career: the problems the dynamical behaviour of comets implied for Cartesian vortex theory.[39] Newton had been doing observations on the comet of 1680 and entered in correspondence about it with the then Astronomer Royal, N. Flamsteed, during the years 1680-1681.

---

[36] G.W. Smith, "The Methodology of the Principia", in: I.B. Cohen and G.W. Smtih, *The Cambridge companion to Newton*, Cambridge University Press, Cambridge, 2002, pp. 138-173.

[37] A. Shapiro, "Newton's Experimental Philosophy", *Early Modern Science and Medicine*, **9**, 3, 2004, pp. 185–217.

[38] The case is discussed *in extenso* in D.T. Whiteside, *Prehistory*, pp. 24-28.

[39] E.g. the "retrograde" trajectories followed by comets with respect to the "natural vortex" around the sun, as well as the impossibility of the observed orientation of a comet's tail when compared to the direction of the motion of its body in a space filled with a liquid aether, especially in the sun's neighbourhood. See Whiteside, *Prehistory*; Ruffner, *Comets*.

But the observations jotted down on that occasion clearly sparked of deeper reflections on the nature of the problems involved. This was a scheme witnessed before. In his student days Newton had already worked on comets:

> *What sparked his interest in things astronomical? The appearance of a comet in late 1664 (...) This he again observed several times some weeks after it was egressed. Afterwards (...) [w]hat underly the visible appearance of the heavens came instead to fill his mind.*[40]

Already in 1664 Newton noticed the problematic nature of cometary motion.[41] His renewed interest in the mathematical properties of planetary motion — fostered by the corrsepondence with Hooke — added to the acuity of his thoughts on observations made on the comet of 1680-1, and his discussion about it with Flamsteed. Indeed, the Ms. produced by Ruffner gives on at least two occasions rise to a profound criticism of Descartes: his theory of light with respect to the visibility and his theory of motion with respect to the orientation of the tail of a comet, especially in the neighbourhood of the sun:

> *The document adds perspective to several areas of Newton Scholarship including his reconsiderations of comets according to the laws of planetary motion following his correspondence with Flamsteed in 1681; his achievements following the correspondence with Hooke in 1679, especially the use of Kepler's*

---

[40] D.T. Whiteside, "The Prehistory of the Principia from 1664 to 1586", *Notes and Records of the Royal Society of London,* **45**, 1, 1991, pp. 11-66. The quote is on p. 17.

[41] Even when his interest shifted over the 1670 to work, apart from alchemy and theology, on light and colours, these problems remain present at the back of his mind. Witness his 1679 attempt to sketch out a (fully Cartesian) lunar theory. Cfr. Whiteside, "Prehistory", p. 17-18.

> *area law; his views on the properties of aether in the early 1680s.[42]*

I think that Newton decided to write out fluid dynamics properly (i.e., in the mathematical way) in order to be able to deal with these problems, but failed. This is what happens in *De Gravitatione*, and it cannot have happened elsewhere than in between 1682 (date of the *de Cometis*) and 1684 (date of *De Motu* 1, in which the new vacuum mechanics is outlined for the first time), because *once* Newton realises that fluid mechanics is not fit to be the sound basis for a system of the world, the next steps follow quickly. The mathematical development of a theory of motion based on the new metaphysics he devised in DG as a consequence of his decisive rejection of Descartes, is resumed from scratch in the manucript known as the *De Motu* 1, probably as a direct consequence of Halley's visit in 1684[43], and which will amount into the first edition of the *Principia*, as explained by Cohen in his *Introduction*.[44]

I shall not analyse in detail the way the definitions (axioms, hypotheses) carry over from DG in the different variants of DM, although that would in itself be a relevant and rewarding exercise. Suffice it to lift out some remarkable points (I base myself on the facsimile edition offered by Whiteside[45]). In the final, often neglected, part of the *De Gravitatione*, Newton for a short while resumes the original hydrostatic project with which the treatise

---

[42] J.A. Ruffner, "Comets", p. 262.
[43] Who up to that moment was still on terms of friendly collaboration with his later enemy Flamsteed, and might as well have heard through him about their correspondence.
[44] I.B. Cohen, *Introduction to Newton's Principia*, Cambridge University Press, Cambdrige, 1971.
[45] D.T. Whiteside, *The preliminary manucripts for Isaac Newton's 1987 Principia 1684-1686*, Cambridge University Press, Cambridge, 1989.

began. He formulates a set of nineteen definitions[46] which deal primarily with the different kinds of forces ("powers"), their "intension" and "extension", and the way they interact with bodies. There is not yet any mention of "vacuum" or "absolute space" (remember we are dealing with an attempt to consistently reformulate the remnants of Cartesian mechanics, as is plain from the comment to def. 15[47]). Now let us turn to the mss. of *De Motu*. There are several documents related to DM 1, the "mothercopy" of which is the tract *De motu corporum in gyrum*. The same is true for DM 2, to which the *De motu corporum liber primus* forms the base. Not only do the (earlier, 1684) tracts of DM 1 all start with sets of definitons relating to kinds of forces and their interaction with bodies, but the attempt by Newton to systematise them (in the tract *De motu corporum in mediis regulariter cedentibus*[48]) follows *exactly* the pattern laid out in de DG, with 19 definitions that express the fundamental concepts.[49] Of course they differ of those in DG by content -, this is the first time that "tempus absolutus", "tempus relativus", "spatium absolutum" and "spatium relativum" appear (defs. 1-4). That they are intended as an explicit revision of the concepts used in DG is indubitably clear from the verso side of the first folio of the DM *in mediis* manuscript, where the summing up of definitions is interrupted, and a list appears that retakes the list of concepts with which DG starts. It is the differences that are revealing:

---

[46] Hall & Hall, pp. 148-152 in their English translation of DG.
[47] *Certainly it suits mathematicians to contemplate things in the light of such reasoning, or if you prefer in the Peripathetic manner; but in physics things seem otherwise.* Hall & Hall, p. 150.
[48] Whiteside, *Preliminary Mss.*, pp. 29-33.
[49] Wrongly numbered, for the last two definitions are both labeled "def. 18". Whiteside, *Preliminary Mss.*, p. 32.

| DG[50]  | DM 1 in med.[51] |
|---------|------------------|
| Locus   | Locus            |
| Corpus  | Quies            |
| Quies   | motus            |
| Motus   | velocitas        |
| -----   | quantitas motus  |

One will appreciate the shift from "corpus" to "quantitas motus" (to which "velocitas" is related), wherein the idea, voiced in the metaphysical part of DG, that body should be defined through the properties required for local motion alone, is at work. The notion of quantity of motion — which *precedes* the notion of absolute space[52] — is essential to the whole conceptual framework of the *Principia*. It is plain that the DM 2 manuscripts are basically but a further elaboration of what was accomplished in DM 1 — after it had been clarified and cleaned up by teaching about it at Cambridge for two years in the fall semesters of 1684 and 1685. Inasfar as the DM manuscripts are preliminary to PR I, DG thus functions as a joint between the 'old' and the 'new' Newton with respect to both physics and metaphysics. That is why I propose to consider DG eventually as the zero release to DM 1 & 2; a DM 0 so to say.[53]

---

[50] Hall & Hall, p. 91.
[51] Whiteside, *Preliminary Mss.*, p. 30.
[52] This important point has been made with admirable clarity by Ori Belkind in his recent paper "Newton's Conceptual Argument for Absolute Space", *International Studies in the Philosphy of Science*, **21**, 3, 2007, pp. 271-293. Nevertheless I do not believe that Belkind's analysis exhausts the whole question, for the argument of the stepwise bigger containers raises problems with respect to infinity (in fact it implies an argument against actual infinity), which I think is difficult to uphold with respect to Newton, for it would prevent him to make the cosmological leap which underpins the universality of the laws governing his system of the world. An inquiry into the consequences of the totalising leap into cosmology can be found in P. Kerzberg, *Critique and Totality*, SUNY Press, Albany, 1997.
[53] I thank Eric Schliesser for the suggestion to call DG a "zero release" to DM.

The small tract "*Elements of Mechanicks*" fits in this picture perfectly well. The structural element of the nineteen definitions is lifted out by the Halls themselves in their commentary.[54] However, they believe it was written after the publication of the *Principia*. They deliver no argument for this except than that this view is more probable than the alternative one, that it was a synopsis of a work Newton intended to write.[55] It will be clear that I do not share their opinion. The structural set-up it has in common with DG and DM 1 points in another direction: that Newton at a certain moment did write up succinctly what he then thought should be the content of PR I. The fact that a question concerning harmonic motion raised by Mersenne has not been retained does not count, for it pertains to the pendulum problem, a subject he did treat extensively in PR I. Revealingly enough the content of the later Book II (the chapter on fluid mechanics) is almost completely absent from this synopsis, while the essence of Book III, the system of the world based on the new mechanics, figures prominently. This again is confirmed by the close relationship between the text on comets and a 1685 manuscript *De Mundi Systemate*[56] that was originally intended to become the second (eventually the third) book of the first *Principia*:

> *The closest parallels are found in the work drafted in early 1685 that he originally intended to be the second book of the Principia, but that was only published posthumously as Newton's System of the World.*[57]

---

[54] Hall & Hall, pp. 85-86.
[55] Hall & Hall, p. 86.
[56] The original title of the ms. being *De motu corporum liber secundus*. The change of the title presumably was the work of J. Conduitt; according to Cohen in the introduction to his facsimile edition of the English translation: I. Newton, *A Treatise of the System of the World*, I.B. Cohen (ed.), Dawsons of Pall Mall, London, 1969, p. xi.
[57] Ruffner, p. 262. The fate of the *System of the World* is in itself interesting, and has been discussed by Cohen in his introduction to his

One will remember that Book III of the *Principia* ends with propositions dealing with the trajectories of comets, and comets are at the heart of the "Mechanical Frame of the world" in the *Elements*.[58] It is rather arbitrary to state that the problems listed towards the end of *Elements* are the loose ends left by PR I (as Hall & Hall do), because several of them have been solved completely, others partially in the course of it. It is therefore safe to place the *Elements* in between DM 1 & 2 and the *Treatise of the System of the World*.

**The first *Principia*: a text in transition**

But PR I is itself in many respects a text still very much in transition, and on matters fundamental, even though its true importance was immediately perceived by Newton's peers.[59] The received story for the origin of the first *Principia* basically runs as follows: it was written on instigation of the then secretary of the Royal Society, Edmund Halley, as a reply to his inquiries into the

---

facsimile edition. The latin original can be found in: S. Horsley (ed.), *Isaaci Newtoni Opera quae extant omnia,* Londini: exc. Joannes Nichols, 1889.
[58] Hall & Hall, pp. 167-168.
[59] Newton's peers perceived the *Principia* from the start as an attempt to replace Descartes, whether they considered the attempt succesful or not. Witness Huygens to Leibniz (in a discussion on *motus verus*: *Je vous diray seulement, que dans vos notes sur des Cartes j'ay remarqué que vous croyez absonum esse nullum dari motum realem, sed tantum relativum. Ce que pourtant je tiens pour tres constant, sans m'arrester au raisonement et experiences de **Newton dans ses Principes de Philosophie**, que je scay estre dans l'erreur, et j'ay envie de voir s'il ne se retractera pas dans la nouvelle edition de ce livre, que doit procurer David Gregorius. Des Cartes n'a pas assez entendu cette matiere.* C. Huygens, *OH* X, n° 2854, p. 614. Huygens's reasons for this, by the editeurs of the *Oeuvres Complètes* (in ft 47, on the same page) rightlfully labelled, "assertion remarquable'' with respect to Newton's *Principia* have to remain undiscussed here, but will be the subject matter of a forthcoming paper.

precise connection between planetary orbits and the inverse square law.[60] I do not want to alter anything to this story's content, only to its interpretation: I say it does not describe the beginning but the end of the gestation of the first *Principia.* Indeed I even believe that Halley's visits to a certain extend led to a premature birth. Not in the sense that the content and mathematical elaboration of the *De Motu*, the theory of motion at the heart of the *Principia's* first Book, leaves anything to be desired. But in the sense that both its methodological implications and its links to the *systema mundi* were not yet thought through all of their consequences entirely. Cohen's unique study on the rôle of hypotheses in Newton's natural philosophy sheds light on the conceptual gap between PR I and PR II & III.[61] I think I can contribute to bridging that gap by placing PR I in a different light. Newton notoriously starts Book III with a set of "Regulae Philosophandi". This set, however, only pops up in the second edition of the *Principia* (1713). In PR I, Book III starts with a series of "Hypotheses", of which only one survives into PR II, while the first two hypotheses change clothes and become the first two regulae. Now the interesting thing is that there is an intermediate phase in the early 1690s when Newton had not yet changed his "Hypotheses" into "Regulae" — a far from innocent adaptation[62] — but did change the content of "Hypothesis III". Cohen discusses the editorial changes[63] made by Newton at the beginning of Book III and says that, to his surprise, Newton added — in the crucial and mentally demanding period 1692-93[64] — during his revision of PR I a Hypothesis IV which he did not himself believe:

> *Newton enlarged the number of "hypotheses" and included a mutually contradictory "Hypoth. III" and "Hypoth. IV", of which the former was a statement of his own position (later printed as "Regula Philosophandi III"), and the latter was decribed as the doctrine of the Aristotelians and the Cartesians."*[65]

---

[60] I.B. Cohen, *Introduction*, 41 sq.
[61] I.B. Cohen, "Hypotheses in Newton's Philosophy", *Physis*, VIII, 1966, pp. 163-184.
[62] *Le changement de terminologie s'eplique, sans doute, par l'aversion croissante de Newton contre les hypothèses ainsi que par un certain glissement dans la signification qu'il attribue à ce terme. Dans la première édtion des Princpia il lui donne le sens traditionel — admission ou supposition fondamentale d'une théorie*. A. Koyré, "Les Regulae Philosophandi", *Archives internationales d'histoire des sciences*, vol. V, 1960, pp. 3-14.
[63] One can consult the *apparatus criticus* to the Cohen-Koyré *variorum* edition of Book III, vol. II, pp. 550-555.

Indeed, who would believe such a thing? The solution is methinks simple however, even if Cohen dismisses it *a priori*. In 1687 Newton either still believed, or at least was still in dubio of Hypothesis IV (III in the first *Principia*). We do know for certain he believed in it up to 1679, for he speaks of his conception of aether related to it in his letter to Boyle as *(...) one conjecture (...) about the cause of gravity. For this end I will suppose* **aether to consist of parts differing** *from one* **another in subtlety by indefinite degrees**.[66] Newton did not only change his mind between the first and the second edition about the meaning of the words "hypothesis" and "phaenomenon", as Cohen points out; he changed as well his position of the cause of gravity. Cohen's assertion about Hypoth. IV is furthermore flatly contradicted by a letter containing the detailed comments Newton had on Huygens's *De la Cause de la Pesanteur* (OH XXI), passed on by Fatio De Duillier[67] to a correspondent:

---

[64] In which period he appears to have been calling many things into question. Cfr. e.g. the letters to Bentley, or the exchanges with Fatio De Duillier concerning the second edition Fatio envisaged to prepare. The Bentley-letters are reprinted in Cohen-Schofield, pp. 271-312 (with an introduction by Perry Miller).
[65] Cohen, "Hypotheses", p. 164.
[66] Letter to Boyle, in I.B. Cohen and R.E. Schofield, *Isaac Newton's Papers and Letters on Natural Philosophy and related documents,* Cambridge University Press, Cambridge, 1958, p. 253.
[67] Fatio De Duillier met Newton at last in 1687, during his first visits to the Royal Society, of which he became a member in 1688. He was in that period a close friend of both Newton and Huygens, as well as an

> *[concerning] Pag. 163 du Traité de Mr Hugens: Monsr Newton est encore indeterminé entre ces deux sentiments. Le premier que la cause de la pesanteur soit inherente dans la matière par un Loi immédiate du Créateur de l'Univers et l'autre que la Pesanteur soit produite par la cause Mechanique que j'en ai trouvée.*[68]

---

aquaintance to Leibniz. He announces the publication of Newton's *Principia* to Huygens three weeks in advance [Fatio to Huygens (24 June 1687; 2465, OH IX, P. 167 sq.]: *Je me suis trouvé trois fois à la Societé roiale ou j'ay entendu proposer tantôt d'assez bonnes choses et tantôt d'assez mediocres. Quelques uns des Messieurs qui la composent sont extrèmement prévenus en faveur d'un livre de Monsr. Newton qui s'imprime presentement et qui se debitera dans trois semaines d'ici. Il m'ont reproché qui je j'étois trop Cartesien et m'ont fait entendre que depuis les meditations de leur auteur toute la Physique étoit bien changée.* These relations are evidenced by a ms. "Ex Epistola cujusdam ad Amicum", which contains a precursor to the claims against Leibniz in *the commercium epistolicum*, in: J. Edleston, *Correspondence of Sir Isaac Newton and Professor Cotes, including letters of other eminent man, now first published from the originals in the Library of Trinity college, Cambrdige; together with an appendix, containing other unpublished letters and papers by Newton*, J.W. Parker, London, 1850, pp. 308-309. On this enigmatic and influential figure there is surprisingly few literature. A recent article fills the lacuna at least to a certain extend: S. Mandelbrote, "The Heterodox Career of Nicolas Fatio De Duillier", in: J. Brooks and I Maclean, *Heterodoxy in Early Modern Science and Religion*, Oxford University Press, Oxford, 2005, pp. 263-296. The 19th century biography by Wollf remains worth reading: R. Wollf, *Fatio de Duillier. Biographien zur Kulturgeschichte der Schweiz 4*, Verlag von Drell, Sükli & Comp., 1862, pp. 67-87.

[68] Fatio à W. De Beyrie, pour Leibniz [1694], n° 2853, OH X, pp. 605-608. This is true beyond question, witness a manuscript in Newton's hand published by the Halls, the draft of a scholium on corr. 4 and 5 of prop. VI, Book III, where we read: *Huius autem generis Hypothesis est unica per quam gravitas explicari potest, eamque geometra ingeniosissimus D.N. Fatio primus excogitavit*. Hall & Hall, p. 313. Fatio's final claim, although relevant, cannot be discussed here, but will be the subject of another paper on Huygens's viewpoint on the nature of gravity, and its influence on Newton. The letter is reprinted in C. Huygens, *Oeuvres Complètes de Christiaan Huygens*, La Société Hollandaise des Sciences, J.A. Vollgraff *et al.* (eds.), 22 vols., Martinus Nijhoff, The Hague, 1888-1950 (OH in what follows). Bopp in his edition of Fatio's *De la Pesanteur* gives in his introduction an overview of the exchange between Newton, Huygens and Fatio in this rspect (see our ftn. 66). Cohen in his Introduction dicsusses the textual material extant concerning the collaboration Newton-Fatio, o.c., pp.

The first "sentiment" is a perfect summary of the idea Newton developed already in DG, and pointing forward to formulations in the *Scholium Generale* to the second and third editions. The second possibility refers to a mechanical model Fatio had developed himself on the basis of Cartesian vortices, but without the fatal Cartesian aequivalence of matter and space. Fatio's model was for a short while being considered as a serious candidate for the causal explanation of gravity on both sides of the Channel, by Newton as well as by Huygens and Leibniz: *With Fatio's prompting, he reconsidered the possibility that some sort of subtle matter or aether might be responsible for the effects of gravity*.[69] During the early 1690s, Newton indeed kept open the possibility that a vacuum is not incompatible with a form of atomistic mechanism in which material dispersion is extremely rare.[70] Thus the vacuum of the scholia to the defintions in Book I, PR I did not at first prevent him to stick to a methodological Hypothesis III of plainly mechanical import.

> *The new Hypothesis III (which became Rule III) was introduced in order to justify the universal gravitation of all bodies in proportion to the quantity of matter which they severally contain and (...) also served to justify discussions concerning the impenetrability*

---

[69] S. Mandelbrote, *Footprints of the lion. Isaac Newton at work*, Cambridge University press, Cambridge, 2001, p. 91.

[70] Cfr. ft 8 to letter n° 2570 (Fatio to Huygens-1690) in OH IX, p. 387. Fatio had written in his own *Pesanteur*, dating from 1691: *Gold contains more void than substance. Water and glass are dense materials, and yet they are almost totally transparent to the passage of light. It is conceivable that all solids could allow almost free passage to sufficiently small particles.* Traces of a like argument can be found in the scholium to corrollary VI in the first edition of the *Principia*. See N. Fatio de Duillier, "De la cause de la Pesanteur", in K. Bopp (ed.), *Drei Untersuchungen zur Geschichte der Mathematik, Schriften der Straburger Wissenschaftlichen Gesellschaft in Heidelberg,* Berlin & Leipzig, pp.19-66, 1929 [1701].

*and the atomic (or, more accurately, particulate) structure of matter.[71]*

Rule III[72] in the second edition of PR methodologically encapsulates the ideas on material qualities Newton had already tentatively formulated in the *De Gravitatione*[73], but apparently only now managed to implement succesfully on the level of his methodology. This complies with the timing Schaffer gives for Newton's "reworking the initial definitions to the *Principia* on transmutation of matter".[74] There are more textual indications to support this interpretation. Gregory mentions around 1706 that he saw a (long awaited!) revised copy for the planned new edition of the *Principia* wherein the "Hypotheses" are changed into "Rules",[75] and we still owe the copy in which these corrections have been made. But the latter ascription only appears in the 1713 second edition of PR.[76]

---

[71] I.B. Cohen, "Hypotheses", p. 175. An intermediate version was copied by Fatio directly from Newton's manuscript and taken to Huygens, who passed it on to Leibniz... *id.,* p. 174.

[72] *Qualitates corporum quae intendi & remitti nequeunt, quaeque corporibus omnibus competunt, in quibus experimenta instituere licet, pro qualitatibus corporum universorum habendae sunt.* A. Koyré and I.B. Cohen (eds.), *Isaac Newton's Philosophiae Naturalis Principia Mathematica, The Third Edition (1726) with variant readings*, Cambridge University Press, Cambridge, 1972 (the "variorum" edition), vol. II, p. 552.

[73] He defines body by means of "sensible abstraction" through the properties required for local motion alone. An apparently crucial point, for he comes back to it: *Thus I have deduced a description of this corporeal nature from our faculty of moving our bodies, so that all the difficulties of the conception may at length be reduced to that* (Hall & Hall, p. 141)*.* (He should add that an analogy between our minds and Divine mind is implicit in his argument.)

[74] Schaffer, *o.c.*, p. 219.

[75] Cohen, *Introduction*, pp. 195-196. Also S. P., Rigaud, *Historical Essay on the First Publication of Sir Isaac Newton's Principia*, Oxford University Press, Oxford, 1838, p. 98 sq.

[76] With an intermediary stage to be found in an addendum to Gregory's *Notae* in which the *new* Hypothesis III (not yet turned into "Regula III"; see Cohen, *Introduction*, p. 190-191) goes with a comment pertaining to the *old* one, stressing the different positions the Catersians and the Peripatetics would take wit regard to it, given their different views on the

With respect to metaphysics, the same pattern arises: only in PR II & III Newton conceptually achieves the final and decisive refutation of Cartesian vortices, and thus the full and complete implementation of the ideas set out already in DG:

> *(...) his main argument against these vortices, put forward forcefully in the second and third edition of both the Principia and the Opticks, emphasized the ineliminability of the inertial effects of the fluid.*[77]

But an interesting intermediary stage is again the (evidently earlier) manuscript which contains the praise for Fatio's ideas on gravity's cause. There we read:

> *For if (...) a certain subtle matter divided into least particles were uniformly scattered through the empty spaces of the heavens and filled as much of their thousandth part; a,d if the bodies of Planets or Comets or other globes were solid and destitute of all pores: these (by Prop XL book II and corol. 2) would lose a thousandth part of their motion (...) And it is reasonable that in a heaven more filled with matter, they would lose a lager part of their motion in proportion to the density, and even more if they were not solid bodies. Whence Comets, which pass through the planetary heavens (...) in all directions would soon lose their motions , and balls of lead shot from guns would very quickly stop unless the spaces of the heaven and of air were nearly vacuous.*[78]

Note the link with the problem of cometary motion. The same

---

unity of matter: *en effet, la troisième, celle justement qui affirmait l'unité de la matière, disparut et fut remplacé par autre chose.* Koyré, o.c., p. 5. Cohen in his comment on Hyp. III as it occurs in PR I does not take this into account; "Hypotheses", p. 171.

[77] G.W. Smith, "Was wrong Newton bad Newton?" in: J.Z. Buchwald, A. Franklin, *Wrong for the Right reasons*, Springer, Dordrecht, 2005, p. 154.

[78] Because of its length, I chose to cite the fragment in the English translation, which I took from Hall & Hall, pp. 315-316.

empirical phaenomena are thus differently explained in different periods of Newton's lifetime. Thus it is only in the second *Principia* that Newton's conceptual evolution reaches its apex, in that the methodological claims and the metaphysical stance are finally made to match completely.[79] The stage is now set for the *Götterdämmerung* with mechanical philosophy as a whole, and especially with its foremost representative, Leibniz, again on the two fronts of methodology (in the priority debate[80]) and metaphysics (in the Leibniz-Clarke correspondence[81]). But that is another, well documented story.

---

[79] This same point — approached from another perspective — is also made in A. Shapiro's article on Newton's experimental philosophy, cited above.

[80] Initiated, not incidentially, by Fatio De Duillier, in his tract on the brachiostome published by the Royal Society in 1699. See N. Guicciardini, *o.c.*, p. 178. The backgrounds of conflict are described in D. Bertoloni Meli, *Equivalence and Priority. Newton versus Leibniz*, Clarendon, Oxford, 1993.

[81] H.G. Alexander, *The Leibniz-Clarke Correspondence: Together wiith Extracts from Newton's Principia and Opticks*, Manchester University Press, Manchester, 1998.

**Conclusion**

> Newton should be understood as fundamentally severing the link between varous speculative — including traditional metaphysical — issues and the development of empirical science.[82]

This is true only to a certain extend. In its full consequences it only becomes reality for the generations after him. It was not even true in the same way throughout Newton's own scientific career. Newton's own intentions remained embedded within an intellectual universe where

> *philosophy, especially metaphysics and epistemology, the physical sciences, and even theology — were interwoven into one overacrhing field called natural philosophy. Hence to treat Newton as a philosopher in a historically accurate way might be to treat him as a natural philosopher.*[83]

I believe I showed in the foregoing that this point of view, taken as a methodological maxim, translates into the requirement to read Newton's conceptual evolution in the area of natural philosophy in close relation to his developments in physics and celestial mechanics, and moreover, that a full appreciation of the one is impossible without an understanding of the other. Given the key rôle Newton's basic concepts played in the origin of what we call to-day modern science, I conclude with the afterthought that we cannot a priori exclude that this fact might continue to reverberate into its realm up to the present.

---

[82] A. Janiak, *Newton as Philosopher*, Cambridge University Press, Cambridge, 2008, p. 42.
[83] A. Janiak, o.c., pp. 8-10.

**Acknowlegments**

Its unexpected short lifespan notwithstanding, the reading group on *De Gravitatione* set up by Eric Schliesser in Leiden during my stay there in 2007-2008 (and which included also Remco van der Geest) proved very fruitful for the development of my ideas concerning it. I furthermore owe a lot to some in depth discussions with Andrew Janiak during his stay in Belgium in april 2008. My understanding of the fundamental issues involved deepened greatly by my attempts to reconcile the insights on Newton's conception of "true motion" and the implications of his cosmological views, respectively developed by Ori Belkind and Pierre Kerzberg. Finally, I had several in depth discussions relating to the different versions of the General Scholium with Steffen Ducheyne. I thank all of them for their generosity in the sharing of their ideas.